\documentclass[pra,twocolumn,showpacs,floatfix,amsmath,amsfonts]{revtex4}\usepackage{epsfig}\usepackage{graphicx}\usepackage{amsmath}\usepackage{amssymb}\usepackage{dcolumn}\begin{document}\title{Solitons of Bose-Fermi mixtures in a strongly elongated trap.} \author{J. Santhanam$^1$}\email{jayanthi@unm.edu}\author{V. M. Kenkre$^1$}\email{kenkre@unm.edu}\author{V. V. Konotop$^2$}\email{konotop@cii.fc.ul.pt}\affiliation{$^1$ Consortium of the Americas for Interdisciplinary Scienceand Department of Physics and Astronomy, University of New Mexico, Albuquerque, New Mexico 87131, USA \\$^2$ Centro de F\'{\i}sica Te\'orica e Computacional,Universidade de Lisboa, Complexo Interdisciplinar, Avenida Professor GamaPinto 2, Lisboa 1649-003, Portugal and Departamento de F\'{\i}sica,Universidade de Lisboa, Campo Grande, Edif'icio C8, Piso 6, Lisboa1749-016, Portugal}\begin{abstract}It is shown that a Bose-Fermi mixture of a degenerate gas of spin-polarized fermions, whose number significantly exceeds the number of bosons, embedded in a strongly anisotropic trap, is described by the one-dimensional coupled nonlinear Schr\"{o}dinger equation for the boson component and the wave equation with external source for the fermion component. Depending on the type of boson-fermion interaction, the system may display modulational instability and the existence of solitons in the fermion and boson components respectively.  Such solitons represent either a local decrease (increase) of the density of both the components or a decrease of the density in one component and an increase of the density in the other component. It is shown that the type of the effective interactions can be easily managed by varying the trap geometry or by means of Feshbach resonance.\end{abstract}\maketitle  \section{Introduction}  Stable coherent structures represent one of the main subjects of interest in the mean-field theory of quantum gases at zerotemperature~\cite{general}. In quasi-one-dimensional condensates of bosonic atoms, such structures are conventionallyassociated with solitons: dark solitons in the case of repulsive interactions among atoms, and bright solitons in the caseof attractive interactions. Successful experimental generation of dark and bright solitons has been reported inBose-Einstein condensates in Refs.~\cite{dark} and Refs~\cite{bright}, respectively. Considerable progress has alsooccurred in the experimental creation of more sophisticated condensates, in particular, boson-fermion(BF) mixtures (see e.g.~\cite{bose-fermi-exper,lithium-exper}). The first theoretical studies of such systems have revealed anumber of interesting properties. One such property is an effective attraction that develops among bosonsdue to the boson-fermion interaction~\cite{TW} which affects the background distribution and the stability properties of thecondensate~\cite{Roth1}. Very recently, numerical work~\cite{Karpiuk} has shown that localized distributions ofcondensed atoms, associated with solitonic trains, can form in a strongly anisotropic trap even when the boson-boson interaction is repulsive. In Ref. \cite{salerno} it has been suggested that the respective localized states, can be interpreted as a matter wave realization of quantum dots.A boson-fermion mixture can exhibit quite different types of behavior according to the relative ratio of the two components. The model explored in~\cite{Karpiuk} and \cite{salerno} addresses a mixture of a relatively small number of fermions embedded in a much larger bosonic component. In this paper, we consider the opposite situation of a relatively smaller number of bosons embedded in a large fermionic component. Such a large fermion component can become degenerate, and only a small portion of the fermions in the vicinity of the Fermi surface participates in the kinetic processes. If the fermions were also spin-polarized, and thus noninteracting, they would represent a linear system. It turns out however~\cite{TW}, that the presence of bosons, even in a relatively small number, can introduce nonlinearity into the fermion system, thereby significantly changing the properties of the system. In the present paper we obtain dynamical equations governing the effectively one-dimensional (1-D) BF mixture, show, in particular, the possibility of the appearance of two-component  (BF) solitons in a strongly anisotropic trap, and describe some of the simplest properties of the solitonic solutions. It is to be noted in this context that two natural limits exists for bosons embedded in a Fermi cloud, depending on the relative densities of the fermions and bosons. One limit is when the boson number dominates the fermion number and is the subject of the analysis in Refs.~\cite{Karpiuk, salerno}. The other limit is when the fermion number dominates the boson number. The latter case, as well as a discussion of the differences in the two cases, is considered in the present paper. The paper is organized as follows. In Section II, we describe the ``Bose-Fermi" system. Section III explains the various length scales of the system. Section IV shows the reduction of coupled 3D equations to a single 1-D equation that governs the system. In Section V we summarize our main results and discuss their significance.\section{Statement of the problem} A mixture of bosons and spin-polarized fermions, with a dominant fermionic component,  can be described by coupled mean-field equations for the boson order parameter $\Psi({\bf r},t)$and the fermion density$n({\bf r},t) = n_{0}(\textbf{r}) + \delta n({\bf r},t)$, where $n_0({\bf r})=\frac{1}{6\pi^2}\left[\frac{2m}{\hbar^2}(E_F-V_F({\bf r}))\right]^{3/2}$, which were derived in~\cite{TW}:\begin{eqnarray}\label{meanfield_b}&& i\hbar\frac{\partial\Psi}{\partial t}+\frac{\hbar^2}{2M} \Delta\Psi -V_B \Psi-g_1|\Psi|^2\Psi-g_2n \Psi=0\\\label{meanfield_f}&&\frac{\partial^2}{\partial t^2}\delta n=\nabla \left[ n_0 \nabla\left( \frac{(6\pi^2)^{2/3}\hbar^{2}}{3m^2n_0^{1/3}}\delta n +\frac{g_2}{m}|\Psi|^2\right)\right]\end{eqnarray}Here $M$ and $m$ are the respective masses of bosons and fermions, the two-body interactions are characterized by the coefficients$g_1=\frac{4\pi\hbar^2a_{bb}}{M},$ and $ g_2=\frac{2\pi\hbar^2a_{bf}}{\mu}$ with scattering cross-sections $a_{bb}$ and $a_{bf}$, and $\mu =\frac{Mm}{M+m}$ (see e.g. \cite{Roth1}), $n_{0}(\textbf{r})$ is the fermion density described by the Thomas-Fermi (TF) approximation for degenerate Fermi gas with Fermi energy $E_{F}$, and $\delta n({\bf r},t)$ is the departure of the  fermion density from $n_0({\bf r})$, considered to be small enough: $|\delta n({\bf r},t)|\ll n_0({\bf r})$. We consider the trap potentials $V$ for both components to be parabolic and strongly elongated along the $x$-direction:$V_B=\frac{M\Omega^2}{2}(\Lambda^2x^2+r^2_\bot)$and $V_F=\frac{m\omega^2}{2}(\lambda^2x^2+r^2_\bot)\,. $Here $ \Lambda$ and $\lambda \ll 1$ in the boson and fermion potentials $V_B$ and $V_{F}$ respectively, represent the high anisotropy, i.e., the respective aspect ratios of the traps: $\Omega$ and $\omega$ are the linear oscillator frequencies of bosons and fermions in the transversedirection, and $r_\bot=(y,z)$. Introducing the dimensionless variables $T=\frac 12 \Omega t$ and ${\bf R}= \frac{{\bf r}}{a}$, where $a=\sqrt{\hbar/M\Omega}$ is the transverse linear oscillator length of the bosons, and re-scaling through\begin{eqnarray}\begin{array}{l}\psi ({\bf R}, T) = 2\sqrt{\pi|a_{bb}|}a \Psi({\bf r},t);\\\rho({\bf R}, T) = 4\pi a^2|a_{bf}| n({\bf r}, t);\end{array}\end{eqnarray}allow the TF distribution for thefermion density $n_{0}({\bf r})$ to be written as\begin{eqnarray}\label{TF}\rho_0= \mathcal{K}_{F}\left[{\cal E}_F-(\lambda^2X^2+R_\bot^2)\right]^{3/2}\end{eqnarray}where $\mathcal{K}_{F} = \frac{2}{3\pi} \frac{|a_{bf}|}{a} \left( \frac{m\omega}{M\Omega}\right)^{3} $ and${\cal E_F}=2\frac{M\Omega}{m\omega}\frac{E_F}{\hbar\omega}$ is the dimensionless Fermi energy. Then Eqs.(\ref{meanfield_b}), (\ref{meanfield_f}) acquire the form\begin{eqnarray}\label{meanfield_b1}i\frac{\partial\psi}{\partial T}&+&\Delta_{{\bf R}}\psi -(\Lambda^2 X^2+R^2_\bot)\psi-2\sigma_1|\psi|^2\psi\nonumber\\&-&\sigma_2\frac{M}{\mu}\rho\psi=0\\\label{meanfield_f1}\frac{\partial^2\rho_1}{\partial T^2}&=&\nabla_{{\bf R}}\cdot\left[\rho_0\nabla_{{\bf R}}\left( \frac{\alpha }{\rho_0^{1/3}}\rho_1+\chi\sigma_2|\psi|^2\right) \right]\label{1.3}\end{eqnarray}where $\rho_1(\bf{R},T)=\rho(\bf{R},T)-\rho_0(\bf{R})$, $\sigma_{1,2}=$sign$(g_{1,2})$, $\alpha =\frac{4}{3} \left[ \frac{3 \pi}{2} \left(\frac{M}{m} \right)^{3} \frac{a}{|a_{bf}|}\right]^{2/3}$ and$\chi=2\left|\frac{a_{bf}}{a_{bb}}\right|\frac{M^2}{\mu m}$.The normalization conditions are\begin{eqnarray}\label{normal2}\int |\psi|^2 d{\bf R}=4\pi\frac{|a_{bb}|}{a}N_b\,\,\,\mbox{and}\,\,\,\int \rho~ d{\bf R}=4\pi\frac{|a_{bf}|}{a}N_f\,,\end{eqnarray}$N_{b}$ and $N_{f}$ being the total boson and fermion numbers.\section{Scaling} In the present paper we consider the case when the transverse dimension of the Fermi component, $R_F$, significantly exceeds the transverse dimension of the Bose component, $R_B$. The orders of the respective scales can be identified as $R_F=\sqrt{{\cal E}_F}$ [which follows from (\ref{TF})] and $R_B=\frac 12$, which in dimensionless units corresponds to the transverse oscillator length of the low density Bose component. Thus we have, ${\cal E}_F \gg 1$.Multiple-scale expansions~\cite{MSPT}, involving a small parameter identified as the ratio between the transverse oscillator length and the healing length, provide a self-consistent and controllable approach in the case of a one-component BEC, and reduce the 3D Gross-Pitaevskii equation to an effective 1D model \cite{BK}. The use of 1D models, in turn, reveal interesting features of the low-dimensional dynamics, allows one to obtain approximate analytical solutions, and results in dramatical reduction of computational resources in studies of the evolution processes. Since BF mixtures are two-component systems characterized by a variety of scales, a natural question arises: Is such a reduction possible in a \emph{Bose-Fermi mixture}? This is the question we address in our present study through a generalization of the approach of ref.~\cite{BK}, when the fermion number significantly exceeds the number of bosons. To this end we first identify the small parameter $\epsilon=\frac{a}{\xi}\ll 1$, where $\xi=(8\pi n_b|a_{bb}|)^{-1/2}$ is the boson healing length ($n_b$ is the maximal boson density). We restrict our analysis to fermion numbers large enough to satisfy $R_F^2/R_B^2\sim {\cal E}_F=\varepsilon_F/\epsilon^2$, where $\varepsilon_F\gtrsim 1$ is a parameter of the problem. Then we can describe the condition ${\cal E}_F \gg 1$, using the same {\em single } small parameter $\epsilon$. This scaling allows us (by analogy with Ref.~\cite{TW}) to concentrate on the spatial domain, limited in the transverse direction by some radius $\tilde{R}$, which satisfies the condition $R_B\ll \tilde{R}\ll R_F$. Physically, such scaling is available due to Pauli's exclusion principle which forces the fermions to occupy a much larger region relative to the bosons. Specifically, we impose $\tilde{R}=\epsilon^{-1/2}$. The elongation of the fermion trap is expressed through $\lambda\lesssim\epsilon^{1/2}$. The spatial domain of the fermions in the longitudinal direction is defined by the constraint, $\epsilon^2\lambda^2X^2\ll 1$, which gives $X\lesssim \epsilon^{-1}$, while for the case of bosons when $\Lambda \sim \lambda \sim \epsilon^{1/2}$, we have that  $X\lesssim \epsilon^{-1/2}$. Thus we consider the situation in which the boson component is embedded inside the fermion system.Subject to the above conditions the TF distribution (\ref{TF}) can be expanded in the Taylor series\begin{eqnarray}\label{approx_rho}\rho_0=\mathcal{K}_{F}{\cal E}_F^{3/2}\left[1-\frac{3\epsilon^2}{2\varepsilon_F}(\lambda^{2} X^2+ R_\bot^2)+ {\cal O}(\epsilon^3)\right].\label{1.3a}\end{eqnarray}We look for a solution in the form of the series$ \psi({\bf R}, T)=\epsilon\psi_1+\epsilon^2\psi_2+\cdots, $ and$\rho({\bf R}, T)=\rho_0+\epsilon^{2} \rho_{1}+\cdots.$To justify the above expansion, we introduce a characteristic width $\ell$ of the bosonic excitation in the axial direction so that the boson density may be  connected to their total number by means of the estimate: $N_b\sim \pi \ell a^2 n$. This allows us to rewrite the small parameter as $\epsilon\sim\sqrt{8N|a_{bb}|/\ell}$. In the case of a localized distribution,  $\ell\sim\xi$, while for the estimates involving plane waves and dark solitons $\ell\sim L$ where $L=a/\sqrt{\Lambda}$ is the longitudinaldimension of the condensate. Then, for excitations with $\ell\sim\xi$, from (\ref{normal2}) we readily obtain the estimate $|\psi_1|^2\sim 1$.On the other hand, the scaling we have introduced implies $|\psi|^2\sim |\rho-\rho_0|$ which corroborates the fact that the boson and fermion numbers involved in the dynamics are of the same order (see also \cite{TW}). Indeed in momentum space, the fermions in question are located in the vicinity of the Fermi surface; the actual number of fermions, $\rho$, is much bigger than its time dependent part $\rho_1$. In other words $|\rho-\rho_0|\ll\rho_0$ is verified.\section{One-dimensional equations}The natural occurrence of the small parameter $\epsilon$ in the problem suggests a perturbation treatment in powers of $\epsilon$. A regular analysis, however, gives rise to secular terms and, consequently, to an inaccurate description. Hence an application of multiple scale analysis becomes necessary. Following standard procedures (see e.g. \cite{BK}),we introduce scaled variables $t_j=\epsilon^jT$ and $x_j=\epsilon^jX$, regarded as independent variables. Substituting the above expansions for time and space into eq.~(\ref{meanfield_b1}) and gathering the terms of the same order in $\epsilon$ we rewrite the equation for a boson macroscopic wave function in (\ref{meanfield_b}), in the form of a set of equations:\begin{eqnarray}\label{order}\left(i\frac{\partial}{\partial t_{0}}+ {\cal L} \right) \psi_{j}=F_j, \end{eqnarray} where $j=1,2,...$, $F_1=0$,\begin{eqnarray*}F_2&=&-i\frac{\partial\psi_{1}}{\partial t_{1}}-2\frac{\partial^{2} \psi_{1}}{\partial x_{0} \partial x_{1}}\\F_3&=&-i\frac{\partial\psi_{1}}{\partial t_{2}}-2\frac{\partial^{2} \psi_{1}}{\partial x_{0} \partial x_{2}} -\frac{\partial^{2} \psi_{1}}{\partial x_{1}^{2}} - i\frac{\partial\psi_{2}}{\partial t_{1}}-2\frac{\partial^{2} \psi_{2}}{\partial x_{0} \partial x_{1}}\nonumber\\&&+ 2\sigma_1 |\psi_{1}|^{2} \psi_{1}+\sigma_2 \frac{M}{\mu}\rho_{1} \psi_{1}\label{F2}\end{eqnarray*}and $j=1,2,...$ coincides with the order of $\epsilon$ at which the equation is obtained. The operator ${\cal L}$ in eq.~(\ref{order}) is given by\begin{eqnarray}\label{L}\begin{array}{l}\displaystyle{{\cal L}\equiv \mathcal{L}_{\bot}+\mathcal{L}_{x} - \sigma_{2}\frac{M}{\mu}\mathcal{K}_{F}{\cal E}_F^{3/2}},\\\displaystyle{\mathcal{L}_{\bot} = - \Delta_{\bot} + \nu^{2}_{\bot} R^2_\bot,\qquad \nu^{2}_{\bot}=1- \sigma_2 \frac{3M}{4\mu} \mathcal{K}_{F}{\cal E}_F^{1/2}}\\\displaystyle{\mathcal{L}_{x} = - \frac{\partial^{2}}{\partial x_{0}^{2}} + \nu^2_{x} x_0^2,\qquad \nu^2_{x}=\Lambda^2- \sigma_{2} \frac{3M}{4\mu}\mathcal{K}_{F}{\cal E}_{F}^{1/2}\lambda^2}\end{array}\label{1.8}\end{eqnarray}Here the respective eigenfunctions of the operators are defined as, $\mathcal{L}_{\bot} \zeta_{lm} = \varepsilon_{lm} \zeta_{lm}({\bf R}_{\bot})$ and$\mathcal{L}_{x} \phi_{n} (x_0) = \eta_{n} \phi_{n}(x_0)$ where $\eta_n=(2n+1)\nu_{x}$, $\varepsilon_{lm}=2(l+m+1)\nu_{\bot}$, and the scaling chosen guarantees that $\nu_\bot>0$.The basic idea of our analysis is to look for a solution to (\ref{meanfield_b1}), which describes the evolution of the background state of the mixture,  in the form:\begin{eqnarray}\psi_{1} = \nu_{x}^{-1/4}A(x_1,t_1) \phi_{0}(x_{0}) \zeta_{0}({\bf R}_{\bot}) e^{-i\omega_0t_{0}}\,.\label{1.10}\end{eqnarray}The envelope function in eq.~(\ref{1.10}) is a function of slow variables, $(x_{1}, x_{2},\cdots,t_{1},t_{2},\cdots)$, from which only the most rapid are indicated explicitly, and is independent of $x_{0},~t_{0}$ and $R_{\bot}$.In eq.~(\ref{1.10}) $\phi_{0}(x_{0})= \left(\frac{\nu_{x}}{\pi}\right)^{1/4}\exp\left(-\frac 12\nu_{x} x_0^2\right)$ and $\zeta_{0}(R_{\bot})=\left(\frac{\nu_{\bot}}{\pi}\right)^{1/ 2}\exp\left(-\frac 12 \nu_{\bot} R_{\bot}^2\right)$ are the normalized eigenfunctions of  ${\cal L}_x$ and ${\cal L}_\bot$, i.e., the ground state wave functions of the 1D and 2-D harmonic oscillators, respectively. The factor $\nu_{x}^{-1/4}$ is introduced to provide the requirement $|A|^2\sim 1$ so that the condition $|\psi_1|^2\sim 1$, established above is satisfied. By direct substitution one verifies that ansatz (\ref{1.10}) satisfies the equation obtained in the first order of$\epsilon$ ($j=1$) when $\omega_0=\sigma_{2}\frac{M}{\mu}\mathcal{K}_{F}{\cal E}_F^{3/2}-(\nu_{x}+2\nu_{\bot})$. The primary goal is the determination of the time evolution of the envelope function, $A(x_1,t_1)$ as a solvability condition \cite{MSPT} for the multiple scales analysis.  To this end we pass to the second order ($j=2$) equation, and find that \begin{eqnarray}\left(i\frac{\partial}{\partial t_{0}}+ {\cal L} \right) \psi_{2}= -i\frac{\partial\psi_{1}}{\partial t_{1}}-2\frac{\partial^{2} \psi_{1}}{\partial x_{0} \partial x_{1}}.\end{eqnarray}Using an ansatz for $\psi_{2}$, such that it is given as the expansion over the complete set of the eigenfunctions $\phi_n(x), \zeta_{m}(R_{\bot})$ and subjecting it to the intial condition, that there are no excited modes $\zeta_{m}(R_{\bot})$ with $m>0$, $\psi_{2}$ can be written as, \begin{equation}	\psi_{2} = \nu_{x}^{-1/4} \sum_{n} B_{n} \phi_{n} \zeta_{0} \exp(-i\omega_{0} t_{0}),\end{equation}Substituting the above form of $\psi_{2}$ into the second order equation (\ref{order}), we find,\begin{equation}\sum_{n} (\omega_{0} - \omega_{n}) B_{n} \phi_{n} \zeta_{0} = \left(-i \frac{\partial A}{\partial t_{1}} \phi_{0}-2\frac{\partial A}{\partial x_{1}} \frac{\partial \phi_{0}}{\partial x_{0}} \right) \zeta_{0}.\nonumber\end{equation}We can see from the above condition that the amplitude $B_{n}$ becomes infinite, when $n = 0$. Such terms are referred to as ``secular terms" \cite{MSPT}. The method of multiple scales allows us to identify and eliminate such secular terms by requiring $F_{2}$ to be orthogonal to $\psi_{1}$. From this condition, we can obtain the solvability condition for $A$ to be $\partial A/\partial t_{1} = 0$ and thus $A$ does not depend on $t_1$. We write $A=A(t_2,x_1)$ and when $n\neq 0$, we find the solution to the second order, \begin{eqnarray}\psi_{2} = \nu_{x}^{-1/4}\frac{\partial A}{\partial x_{1}} \sum_{n =1 }^\infty \frac{\Gamma_{n,0}}{\eta_{0} - \eta_{n}} \phi_{n}(x_0)\zeta_{0}({\bf R}_\bot) e^{-i\omega_{0} t_{0}}.\label{1.11a}\end{eqnarray}where  $\Gamma_{n,0}=-2\int_{- \infty}^{\infty}\phi_{n}(x)\frac{d}{dx}\phi_{0}(x)dx$.The equation of the third order, however, involves the dependence on the fermion density, which means that it must be solvedtogether with Eq. (\ref{meanfield_f1}), which in the new, scaled variables, after substituting for $\psi_{1}$ from (\ref{1.10}), acquires the form\begin{eqnarray}\frac{\partial^{2} \rho_{1}}{\partial t_{0}^{2}} =v_{F}^{2}\nabla_{R_{0}}^{2} \rho_{1} + \chi^{'}(R_{0},t_{0}) |A|^{2},\label{1.14b}\end{eqnarray}where $R_{0} = (x_{0},y,z)$ and $v_{F}^{2} = \alpha {\cal K}_{F}^{2/3} {\cal E}_F $.\begin{eqnarray}\label{chi}\chi^{\prime} = \chi \mathcal{K}_{F} {\cal E}_F^{3/2} \sigma_{2} \left(\phi_{0}^{2}\nabla_{\bot}^{2} \zeta_{0}^{2} +|\zeta_{0}|^{2}\frac{\partial^{2}}{\partial x_{0}^{2}}\phi_{0}^{2} \right)\,.\end{eqnarray}In the physical variables, ${\rm v}_F^2=\frac{1}{4}v_{F}^{2}\Omega^2a^2=\frac{4}{3}\frac{E_F}{2m}$,and therefore in what follows, we refer to $v_F$ as the Fermi velocity. Now Eq. (\ref{1.14b}) acquires a transparent physical meaning. Fermions as quasi-particles are excited on the Fermi surface, and thus propagate with the Fermi velocity along the condensate. Their dynamics is described by the wave equation with a source term, which describes the effect of the boson distribution on the fermion density.Because $A(x_{1},t_{2})$ is independent of $R_{\bot},~x_{0}$ and $t_{0}$, it can be treated as a constant in Eq. (\ref{1.14b}). Since $\phi_{0}(x_{0})$ changeson the scale of the longitudinal dimension of the boson component, $\ell\sim\xi$, while $\zeta_{0}(R_{\bot})$ changes on the scale of transversedimension $a$, the effect defined by the curvature of $\zeta_{0}^{2}(R_{\bot})$ dominates in  $\chi^{\prime}$and we can neglect the second term in Eq.~(\ref{chi}). We are interested in coupled fermion-boson excitations (in particular, solitons) when $\rho_1$ is represented by the particular solution of the in homogeneous equation (\ref{1.14b}) as shown below: \begin{eqnarray}\label{fdensity}\rho_{1} =&&-\frac{\chi v_{F}\sigma_{2}}{\pi^{3/2}\alpha^{3/2}}\nu_{\bot}|A|^{2}e^{-\nu_{\bot} R_{\bot}^{2}}.\label{1.15}\end{eqnarray}Comparing $\rho_{1}$ to $\psi_{1}$ from eq.~(\ref{1.10}), given by\begin{equation}\label{psi1}\psi_{1} = \sqrt{\frac{\nu_{\bot}}{\pi}}A(x_1,t_2) e^{-\frac{\nu_{x}}{2} x_0^2} e^{-\frac{\nu_{\bot}}{2} R_{\bot}^{2}} e^{-i\omega_0t_{0}}, \end{equation}we find that $\rho_{1}$ is also governed by the same envelope function $A(x_{1},t_{2})$ as the bosons. Substituting this expression for $\rho_{1}$ in $F_3$ we can write the third order expansion $(j=3)$ for the bosons as \begin{eqnarray}	\left(i\frac{\partial}{\partial t_{0}}+ {\cal L} \right) \psi_{3}= -i\frac{\partial\psi_{1}}{\partial t_{2}}-2\frac{\partial^{2} \psi_{1}}{\partial x_{0} \partial x_{2}} -\frac{\partial^{2} \psi_{1}}{\partial x_{1}^{2}} \nonumber\\- i\frac{\partial\psi_{2}}{\partial t_{1}}-2\frac{\partial^{2} \psi_{2}}{\partial x_{0} \partial x_{1}} + 2\sigma_1 |\psi_{1}|^{2} \psi_{1}\nonumber\\-\left( \frac{\chi v_{F}\nu_{\bot}}{(\pi\alpha)^{3/2}}\frac{M }{\mu}\right)|A|^{2}e^{-\nu_{\bot} R_{\bot}^{2}} \psi_{1}.\label{boson3}	\end{eqnarray}Here we have used the fact that $\sigma_{2}^{2} =1$. Requiring the absence of secular terms in the third order equation of the multiple scale expansion, which is the same as requiring the orthogonality of $F_3$ with $\psi_2$ and $\psi_1$, and employing explict expressions for $\psi_1$ and $\psi_2$, given by (\ref{psi1}) and (\ref{1.11a}), respectively,we find that $A(x_{1},t_{2})$, satisfies,\begin{eqnarray}i\frac{\partial A}{\partial t_{2}} + D \frac{\partial^{2} A}{\partial x_{1}^{2}} + \chi_{eff} |A|^{2}A = 0,\label{B-Cf}\end{eqnarray}which is the solvability condition we seek. In eq. (\ref{B-Cf}), $D = 1 + \sum_{n =1}^\infty \frac{|\Gamma_{n0}|^{2}}{\eta_{0}-\eta_{n}} = 1/2$; $\chi_{eff}=-\chi_{bb}+\chi_{bf}$, is the effective nonlinearity characterizing the combined effect of: direct boson-boson interactions, $\chi_{bb} = \frac{2\sigma_{1}}{(2\pi)^{3/2}}\nu_{\bot}$, and the interactions among bosons induced by boson-fermion scattering processes,  $\chi_{bf}=\frac{\chi M v_{F}}{2\pi^{3/2}\alpha^{3/2}\mu}\nu_{\bot}$. The solution to equation (\ref{B-Cf}) is known exactly. This is the main result of our analysis and shows that bright and dark soliton solutions are possible for the envelope function. According to the method of multiple scales, $t_2=\epsilon^{2} T$ and $x_1 =\epsilon X$, thus we can re-write the solution, $A(x_{1},t_{2})$ of the eq.~(\ref{B-Cf}) in terms of our original scaled variables, $X$ and $T$, as $A(\epsilon X,\epsilon^{2}T)$. Eq.~(\ref{B-Cf}) along with eqs.~(\ref{psi1}) and (\ref{fdensity}) describe the evolution of the original system as defined by (\ref{meanfield_b1}) and (\ref{meanfield_f1}) for times up to $1/\epsilon^2$ ($\epsilon \ll 1$).When the boson-fermion interaction is turned off ($\chi_{bf} = 0$), we expect the boson and fermion components to be independent of each other, i.e., eqs.~(\ref{meanfield_b1}) and (\ref{meanfield_f1}) are decoupled. In such a case, $\chi_{eff}=-\chi_{bb}$, and from eqs. (\ref{1.15}), (\ref{psi1}) and (\ref{B-Cf}), we see that the fermion density is given by the Thomas-Fermi Approximation, eq. (\ref{TF}) and the boson system is described by NLS equation as expected. When $\chi_{bf} \neq 0$, we see that boson-fermion interactions always lead to attractive (effective) interactions between the bosons. This qualitatively agrees with earlier results~\cite{TW}. Our analysis below shows that the quantitative dependence of the behavior of the mixture on the system parameters is very different for the case when the number of fermions is much less than the number of bosons compared with the case when the number of fermions is much larger than the number of bosons. \section{Discussion and estimates for real condensates} Since our analysis has resulted in the NLS equation (\ref{B-Cf}) governing the dynamicsof small amplitude excitation in BF mixtures, we know at once that several well-known consequences of the NLS equation will follow, including {\em modulational instability} of plane waves, and existence of {\em bright} and {\em dark} solitons, depending on the sign of $\chi_{eff}$. When $D\chi_{eff} > 0$, the envelope function, $A$ is given by the bright soliton solution ($``\rm{sech}(\epsilon X)"$) and when $D\chi_{eff} < 0$, it is given by the dark soliton solution ($``\tanh(\epsilon X)"$). Since $D > 0$, the bright or dark soliton solution is solely dictated by the sign of $\chi_{eff}$. In the case where boson-boson and boson-fermion interactions  have different signs, the observable phenomena depends on the relative values of $\chi_{bb}$ and $\chi_{bf}$. One can easily find the critical value for the boson-fermion interaction at which both interactions are balanced, i.e. when $\chi_{eff} = 0$:\begin{equation}|a_{bf}^{cr}| = \left[ \pi \frac{\mu^{2}}{m^{2}}\sqrt{\frac{m}{M}\frac{\hbar \Omega}{E_{F}}}\right]^{1/2}\sqrt{a\,a_{bb}}\label{critical}\end{equation}This value that we obtain differs from that reported in \cite{Karpiuk}. This is expected given that the underlying physical conditions are different in the two cases. Note, in particular, the weak dependence of $a_{bf}^{cr}$ in eq. (\ref{critical}) on the fermion number, $|a_{bf}^{cr}|\sim N_F^{-1/12}$ in contrast to the dependence $|a_{bf}^{cr}|\sim N_F^{-1/2}$ reported in \cite{Karpiuk}. This is due to the fact that in our case we consider a small number of bosons interacting with a large system of degenerate Fermi liquid and changing the number of fermions does not have much impact while in \cite{Karpiuk} the small number of fermions are trapped by a large system of bosons. In such a case, the effect of interaction with each fermion becomes important and hence a large dependence on the fermion number. In our result, we also find that $|a_{bf}^{cr}|\sim \sqrt{a_{bb}}$ while in~\cite{Karpiuk} $|a_{bf}^{cr}|\sim a_{bb}^{2/5}$.When $|a_{bf}|$ equals the critical value, the system becomes an ideal gas, described by the linear Schr\"{o}dinger equation.If $|a_{bf}|$ is larger than the above critical value, $\chi_{bf} > \chi_{bb}$, interaction between bosons and fermions dominates,leading to a negative effective attractive interaction between the bosons. This is the case where modulational instability and bright solitons can be observed. Eq.~(\ref{critical}) shows that value of $|a_{bf}^{cr}|$ can be manipulated by changing the radial trap size, $a$.Note that using the parameters of the Rb$^{87}$-K$^{40}$ mixture experiment~\cite{bose-fermi-exper}: $\Omega = 2\pi\times 197$ Hz, $a_{bb} = +5.25$ nm,and the Fermi temperature $0.25~\mu$K, allow us to compute $|a_{bf}^{cr}|\approx 32$ nm. On the other hand, using the data of Li$^{7}$-Li$^{6}$ mixture, Ref.~\cite{lithium-exper}: $\Omega = 16.14$ kHz, $a_{bb} = -1.35$ nm,  the Fermi temperature $4~ \mu$K, gives us $|a_{bf}^{cr}|\approx 2.32$ nm. An alternative way of making the boson-fermion interactions to be dominant is to increase $a_{bf}$ (or decrease $|a_{bb}|$)\cite{Karpiuk}.\begin{figure}[h!]\includegraphics[height= 5.7cm, width = 8cm]{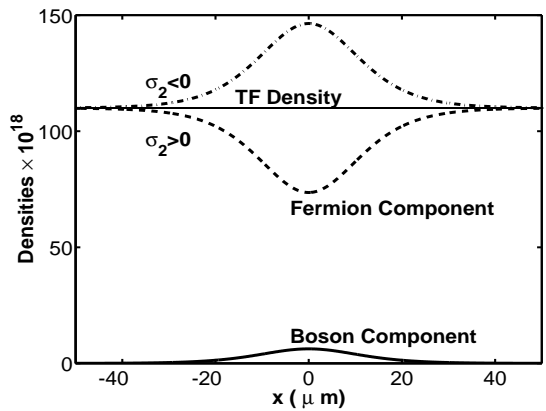}\caption{Two-component soliton formation along the elongated trap direction ($x$-direction) for the Rb$^{87}$-K$^{40}$ mixture example. The solid line shows the density of bosons, $|\Psi(\textbf{r},t)|^{2}$, corresponding to the bright soliton solution for the envelope function. The dash and dash-dot lines shows the fermion density, $\delta n(\textbf{r},t)$, for repulsive and attractive interactions between bosons and fermions, respectively. Taking into account the nonzero background, the excitation of fermions in the former case can be associated with a conventional dark soliton, while in the later case, it is a bright-against-a-background soliton.}\label{1.1}\end{figure}An interesting peculiarity of the bright soliton solution of eq.~(\ref{B-Cf}), follows from (\ref{1.15}). If the fermion-boson interaction is repulsive($\sigma_2>0$), the  fermion density, has a relative negative sign compared to the boson density and is lower in the region of the soliton than outside. In thiscontext, the terminology ``bright soliton'' does not reflect the situation precisely, because in the fermionic component oneobserves darkening. To illustrate this, we plot the densities of bosons (solid line) and fermions for the  for $\sigma_2<0$ (dash-dot line) and $\sigma_2>0$ (dashed line) along the elongated trap direction ($x$-direction) in Fig. \ref{1.1} for a Rb$^{87}$-K$^{40}$ mixture \cite{bose-fermi-exper}, when the envelope solution of eq.~(\ref{B-Cf}) is a bright soliton ($\chi_{eff} >0$). While for attractive boson-fermion interactions one haseither bright-bright or dark-dark solitons, in the case of repulsive boson-fermion interactions, one can excite bright-dark anddark-bright solitons. Here, the first and second term refer to bosons and fermions respectively (see also Fig.~\ref{1.1} for the explanation of the terminology for the fermions). For the calculations presented in the present paper, the choice of an appropriate scaling in finding the small parameter $\epsilon$, was of primary importance. In order to check its feasibility for experimentally available atomic gases, we take into account the possibility of a change of the scattering length by meansof Feshbach resonance and consider an example using Rb$^{87}$-K$^{40}$ mixture~\cite{bose-fermi-exper}, with a radial size of$a \approx 2~ \mu$m, and longitudinal extension $L\approx 200~\mu$m with the Fermi temperature of order of 0.25 $\mu$K. Takingthe number of rubidium atoms to be $500$, the scattering length $a_{bb} = 0.1$ nm,  and the width of the soliton $\ell\approx 20~\mu$m, we obtain the mean healing length $\xi\approx 14\,\mu$m, which gives $\epsilon = 0.14$,${\cal E}_{F} = 3910 \gg 1$ and $\varepsilon_F\approx 76.6$. We also notice that for the case of the Li$^7$-Li$^6$ mixture~\cite{lithium-exper} with the Fermi temperature $4~\mu$K we obtain ${\cal E}_{F} = 500 \gg 1$, which also matches with the conditions of the application of the theory. Fig~1 shows the formation of the two-component soliton for such a Rb$^{87}$-K$^{40}$ mixture with $|a_{bf}| = 20$ nm, when the boson-fermion interaction is attractive ($\sigma_{2}<0$) andrepulsive ($\sigma_{2}>0$).\section{Conclusion} In this paper we have considered a boson-fermion mixture, where the number of fermions is much larger than the number of bosons and is confined by a strongly anisotropic trap. It has been shown that both the boson and fermion systems are described by an effective 1D nonlinear Schr\"{o}dinger equation (\ref{B-Cf}) and thus can display modulational instability and the existence of bright and dark solitons. Such solitons involve most of the bosons of the system and a relatively small portion of the fermions in the vicinity of the Fermi surface, where the soliton propagation is along the axial direction of the condensate. In a system without coupling between the bosons and the fermions, the fermions would tend to propagate with their Fermi velocity. However, the coupling to the bosons that is present in our system forces them to form a slow (and even static) solitary excitation following the behavior of the bosons. A remarkable feature of the result obtained is that the soliton dynamics is governed by a single nonlinear Schr\"{o}dinger (NLS) equation, even though, physically, it describes two separate systems. The soliton solution of the NLS equation can describe either an increase or a decrease in the densities of the two components. In the case of the attractive interaction between the components, the solitons may describe simultaneous increase or decrease of the densities of both the components of the mixture. In the case of the repulsive interactions between the components on the other hand, they can describe an increase of the density in one component and a decrease of the density in the other component. This is quite an unusual situation: there is only one soliton solution to the NLS equation for a given effective nonlinearity, however it describes opposite effects in the two components. Another feature to be mentioned is that due to the finite extension of the condensate in the longitudinal direction, an internal mode is excited. This excitation results in the effective mass of the bosons, making up the soliton, being twice as large as the real mass, which does not happen in the case of infinite extent of the trap. We have explicitly shown that the type of the effective interactions can be experimentally manipulated either by just affecting the trap geometry or by means of Feshbach resonance (the latter option being also efficient in the opposite case of relatively small number of fermions~\cite{Karpiuk}). We hope that observational attempts to verify these physical aspects of the results we have obtained will be made in future experiments.\acknowledgements We thank Francisco Sevilla for discussions. V.V.K. acknowledges Takeya Tsurumi for sending a reprint of the paper~\cite{TW}. V.M.K. acknowledges partial support of DARPA under DARPA-N00014-03-1-0900 and the NSF under INT-0336343.   The work of V.V.K was supproted by the FCT and european program FEDER under the grant POCI/FIS/56237/2004.
\begin{thebibliography}{99}\bibitem{general} see e.g. L.~P.~Pitaevskii, and S.~Stringari, {\em Bose-Einstein Condensation} (Oxford University Press, Oxford,2003).\bibitem{dark} M.~R.~Andrews, D.~M.~Kurn, H.-J.~Miesner, D.~S.~Durfee, C.~G. Townsend, S.~Inouye, and W.~Ketterle, Phys. Rev. Lett. {\bf 79}, 553 (1997); S.~Burger, K.~Bongs, S.~Dettmer, W.~Ertmer,  K.~Sengstock, A.~Sanpera, G.~V.~Shlyapnikov, and M.~Lewenstein, Phys.  Rev. Lett. {\bf 83}, 5198 (1999); J. Denschlag, J. E. Simsarian, D. L. Feder, Charles W. Clark, L. A. Collins, J. Cubizolles, L. Deng, E. W. Hagley, K. Helmerson, W. P. Reinhardt, S. L. Rolston, B. I. Schneider, W. D. Phillips, {\it Science} {\bf 287}, 97 (2000).\bibitem{bright}K.E. Strecker, G.B. Partridge, A.G. Truscott and R.G. Hulet, {\it Nature} {\bf 417}, 150 (2002);L. Khaykovich, F. Schreck, G. Ferrari, T. Bourdel, J. Cubizolles, L.D. Carr, Y. Castin and C. Salomon, {\it Science} {\bf 296}, 1290 (2002). \bibitem{bose-fermi-exper}G. Modugno, G. Roati, F. Riboli, F. Ferlaino, R. J. Brecha, M. Inguscio,  Science {\bf 297}, 2240 (2002).\bibitem{lithium-exper}F. Schreck, G. Ferrari, K. L. Corwin, J. Cubizolles, L. Khaykovich, M.-O. Mewes, and C. Salomon, Phys.\ Rev. A {\bf 64}, 011402(R) (2001).\bibitem{TW}T. Tsurumi and M. Wadati J. Phys. Soc. Jap. {\bf 69} 97 (2000)\bibitem{Roth1}R. Roth, Phys. Rev. A {\bf 66}, 013614 (2002)\bibitem{Karpiuk}T. Karpiuk, M. Brewczyk, S. Ospelkaus-Schwarzer, K. Bongs, M. Gajda and K. Rz\c{a}\.{z}ewski Phys. Rev. Lett. {\bf 93}, 100401 (2004)\bibitem{salerno} M. Salerno, cond-mat/0503097.\bibitem{MSPT} see e.g.J.~Kevorkian and J.~D.~Cole, {\em Multiple Scale and Singular Perturbation Methods} (Springer Press, New York, 1996), Chapter 6.\bibitem{BK}V. A. Brazhnyi and V. V. Konotop, Mod. Phys. Lett. B {\bf 18}, 627 (2004).\end{thebibliography}
\end{document}